# Существование, единственность и монотонность решения задачи потокораспределения в гидравлических цепях с зависящими от давления замыкающими соотношениями

Корельштейн Л.Б. (ООО «НТП Трубопровод»)

В статье доказывается существование и единственность решения классической задачи потокораспределения в гидравлических цепях с ветвями, имеющими зависящие от давления характеристики (замыкающие соотношения). При этом вид характеристик ветвей ограничен только условиями, необходимыми для обеспечения существования и единственности решения задачи потокораспределения для каждой отдельной ветви (непрерывность и монотонность), а также непрерывности такого решения. Показано, что при этом гидравлическая цепь в целом «наследует» монотонность и непрерывность поведения своих ветвей, что и обеспечивает существование и единственность решения. Установлены также некоторые свойства матрицы Максвелла гидравлической цепи, связанные со свойствами монотонности решения.

Вопрос о существовании и единственности решения задачи потокораспределения в гидравлических цепях всегда вызывал интерес исследователей, занимающихся расчетами трубопроводных систем. В его исследование внесли вклад как отечественные (Б.Н. Пшеничный [1]), так и зарубежные (M. Collings и др. [2]) специалисты. Этот вопрос всесторонне рассматривался основоположниками теории гидравлических цепей А.П. Меренковым и В.Я. Хасилевым [3], и М.Г. Сухаревым [4, 5]. Последние результаты по этому вопросу были получены С.П. Епифановым и В.И. Зоркальцевым, которым удалось установить существование и единственность решения при наиболее общих предположениях о характеристиках ветвей как для классической задачи потокораспределения [6], так для целого ряда расширенных или неклассических задач [7, 8, 9, 10]. Во всех перечисленных выше случаях был использован экстремальный подход - задача сводилась к поиску минимума строго выпуклой функции, что и обеспечивало существование и единственность решения.

Тем не менее все перечисленные выше результаты получены в предположении, что расходы по ветвям цепи зависят только от разности потенциалов (давлений) на концах ветвей, но не от величины самих давлений. Такое предположение вполне правомерно для течения несжимаемой жидкости при изотермических условиях, однако во многих случаях не применимо при течении реальных газов и тем более двухфазных газо-жидкостных потоков при изотермических или адиабатических условиях, когда теплофизические свойства продукта сильно зависят от давления. При этом для гидравлических цепей с зависящими от давления замыкающими соотношениями, как отмечено в статье М.Г. Сухарева [5], использовать экстремальный поход (по крайней мере пока) не удается. Тем не менее для таких сетей, как правило, можно эффективно использовать обобщения классических методов решения для «традиционных» гидравлических сетей (такие обобщения предложены, в частности, в работах Е.А. Михайловского и Н.Н. Новицкого [11, 12, 13]). Однако отсутствие строго математически установленных условий существования и единственности решения оставляло (по крайней мере у автора) чувство неудовлетворенности – тем более что уже давно известно, что в весьма схожих задачах неизотермического течения, в особенности двухфазных продуктов, возможны случаи не единственности решения!

В данной работе содержится доказательство существования и единственности решения классической задачи потокораспределения для подобных цепей, причем при самых общих ограничениях на характеристики ветвей, прямо обобщающих аналогичные ограничения в работах С.П. Епифанова и В.И. Зоркальцева [6-10].

# 1. Постановка задачи и ограничения на характеристики ветвей

Пусть $G$ - ориентированный граф с $N_V$ узлами (образующими множество узлов $V$) и $N_E$ ветвями (образующими множество ветвей $E$). Расход $X_i$ по i-той ветви связан с начальным и конечным давлениями $P_{Fi}$ и $P_{Li}$ замыкающим соотношением

$$X_i = \varphi_i(P_{Fi}, P_{Li}) \qquad (1)$$

Пусть $A$ - матрица инцидентности графа $G$ ($a_{ij} = 1$, если ребро j начинается в узле i; $a_{ij} = -1$, если ребро j заканчивается в узле i; $a_{ij} = 0$ в остальных случаях); $Q$ - вектор узловых притоков. Тогда уравнения Кирхгофа (уравнения балансов в узлах) записываются в виде

$$AX = Q \qquad (2)$$

Используя матрицы $A_F$ и $A_L$, соответствующие выходящим и входящим ветвям ($A = A_F + A_L$), вектор узловых давлений $P$ и вектор $\Phi$ функций $\varphi_i$, уравнения (1) можно записать в виде

$$X = \Phi(P_F, P_L), \ P_F = A_F^T P, \ P_L = -A_L^T P \qquad (3)$$

Таким образом, имеем $N_V + N_E$ уравнений для $2N_V + N_E$ неизвестных ($P$, $Q$ и $X$). При этом, как известно, уравнения (2) не являются независимыми – для связного графа $G$ матрица $A$ имеет ранг $N_V - 1$, при этом притоки в узлах удовлетворяют дополнительному уравнению баланса:

$$\sum_{i=1}^{N_V} Q_i = 0 \qquad (4)$$

Как видим, для связного графа $G$ неизвестных на $N_V$ больше, чем независимых уравнений, поэтому, чтобы задача была определенной, должны быть заданы значения $N_V$ неизвестных.

В классической задаче потокораспределения (КЗП) задается вектор давлений $P_{fix}$ в $N_P > 0$ узлах (образующих множество $V_P$) и вектор притоков $Q_{fix}$ в остальных $N_Q = N_V - N_P$ узлах (образующих множество $V_Q$), при этом требуется найти давления $P_{var}$ в остальных $N_Q$ узлах, расходы по ветвям $X$ и притоки $Q_{var}$ в $N_P$ узлах с заданным давлением. Последние определяются по уравнениям (1) и (2), так что найти, по существу, достаточно давления $P_{var}$ в узлах с заданными притоками.

Для «традиционных» гидравлических цепей функции $\varphi_i$ зависят только от разности давлений: $X_i = \varphi_i(P_{Fi} - P_{Li})$. В работе [6] сформулированы условия для функций $\varphi_i$ (или обратных к ним функций $f_i$), при которых решение КЗП для «традиционных» гидравлических цепей гарантированно существует и единственно, а именно (Условия A):

1) Непрерывность;
2) Строгое монотонное возрастание;
3) Определенность на всем множестве действительных чисел $\mathbb{R}$;
4) Совпадение области значений с $\mathbb{R}$, что с учетом монотонности эквивалентно условиям $\varphi_i(y) \to +\infty$ при $y \to +\infty$ и $\varphi_i(y) \to -\infty$ при $y \to -\infty$.

Строгая монотонность функции необходима для обеспечения единственности решения; непрерывность и возрастание вытекают из физических соображений. Эти условия почти всегда реализуются на практике.

А вот условия 3) и 4) являются только удобной математической экстраполяцией, призванной обеспечить существование решения при любых заданных давлениях и узловых притоках – на самом деле на практике область значений давлений всегда ограничена – почти всегда снизу (обычно как минимум положительностью абсолютных давлений), а часто и сверху (технологическими ограничениями, прочностью конструкции и т. д.), а соответственно ограничены и величины возможных расходов и узловых притоков. Поэтому (как отмечено и в [9]) на практике желательно иметь оценки границ величин заданных узловых давлений и узловых притоков, при которых задача заведомо имеет решение в рамках заданных границ изменения аргументов функций $\varphi_i$ и $f_i$. Это, однако, самостоятельная (весьма практически важная и непростая) задача.

Множество функций, удовлетворяющих описанным выше условиям 1)-4), обозначим как $\tilde{Z}_a$. Оно отличается от введенного в [6-9] множества $\tilde{Z}$ только отсутствием условия равенства нулю в нуле. Множество функций $\tilde{Z}_a$ получается из $\tilde{Z}$ прибавлением к функциям произвольной постоянной. Функции из $\tilde{Z}$ представляют собой так называемые пассивные ветви (без напора), а функции из $\tilde{Z}_a$ включают в себя и активные ветви (с перепадом высот, насосами, компрессорами и т.п.).

Определим теперь множество допустимых функций $\tilde{Z}_a^2$ для зависящих от давления замыкающих соотношений. Оно представляет собой естественное обобщение множества $\tilde{Z}_a$. Функция $\varphi \in \tilde{Z}_a^2$, если выполняются следующие условия (Условия А2):
1) $\varphi(P_F, P_L)$ является непрерывной функцией двух переменных $P_F, P_L$;
2) $\varphi(P_F, P_L)$ при любом значении $P_F$ строго убывает по $P_L$;
3) $\varphi(P_F, P_L)$ при любом значении $P_L$ строго возрастает по $P_F$;
4) $\varphi(P_F, P_L)$ определена на всем $\mathbb{R}^2$.
5) При любом $P_F$ $\varphi(P_F, P_L) \to -\infty$ при $P_L \to +\infty$ и $\varphi(P_F, P_L) \to +\infty$ при $P_L \to -\infty$
6) При любом $P_L$ $\varphi(P_F, P_L) \to +\infty$ при $P_F \to +\infty$ и $\varphi(P_F, P_L) \to -\infty$ при $P_F \to -\infty$

Иначе говоря, множество $\tilde{Z}_a^2$ составляют непрерывные функции, которые при любом $P_L$ принадлежат $\tilde{Z}_a$ как функции от $P_F$ и при любом $P_F$ принадлежат $-\tilde{Z}_a$ как функции от $P_L$.

В [11-13] приведен ряд примеров принадлежащих множеству $\tilde{Z}_a^2$ функций (описывающих характеристики газоперекачивающих агрегатов).

Так же, как и в случае множества $\tilde{Z}_a$, условия 1)-3) основываются на физических соображениях – а условия 4)-6) являются математической экстраполяцией.

Определим подмножество функций $\tilde{Z}^2$ множества $\tilde{Z}_a^2$, как функции, для которых $\varphi(P,P) = 0$ при любом $P$. Функции из $\tilde{Z}^2$ представляют собой характеристики пассивных ветвей, а $\tilde{Z}_a^2$ включают и активные ветви.

Заметим, что любая зависящая только от разности давлений характеристика ветви, удовлетворяющая какому-либо условию из списка А, удовлетворяет и соответствующему условию(ям) из списка А2. Тем самым все установленные далее результаты будут справедливы и для «традиционных» гидравлических цепей.

## 2. Вспомогательные определения и свойства

Прежде всего получим некоторые вспомогательные результаты, которые понадобятся для дальнейшего изложения.

Установим некоторые свойства функций из множества $\tilde{Z}_a^2$.

Пусть $\varphi \in \tilde{Z}_a^2$. Тогда уравнения $\varphi(P_F, P) - X = 0$ и $\varphi(P, P_L) - X = 0$ относительно $P$ определяют неявные функции $f_L(P_F, X)$ и $f_F(P_L, X)$, которые рассчитывают давление в конце ветви по давлению в начале и расходу, и наоборот, давление в начале ветви по давлению в конце и расходу. Легко видеть, что в силу определения $\tilde{Z}_a^2$ эти функции определены на всем множестве $\mathbb{R} \times \mathbb{R}$, монотонны по обоим параметрам (первая монотонно возрастает по давлению в начале ветви и монотонно убывает по расходу; вторая монотонно возрастает по обеим аргументам), и стремятся к бесконечности при стремлении к бесконечности любого из аргументов. Менее тривиальным фактом является то, что эти функции также непрерывны по **обеим** переменным. Последнее вытекает из строгой монотонности функций $\varphi(P_F, P) - X$ и $(P, P_L) - X$ по $P$, и специального «недифференциального» варианта теоремы о неявной функции, предложенного K. Jittorntrum [14] и окончательно доказанного S. Kumagai [15] (формулировку данного варианта см. в Приложении 1).

Заметим также, что направление любого ребра в нашей постановке задачи задано исключительно для удобства записи уравнений (чтобы условиться, в каком направлении расход считается положительным). Для удобства направление ребра всегда можно поменять, определив замыкающее соотношение для ребра с обращенным направлением $\varphi^*(P_F, P_L) = -\varphi(P_L, P_F)$.

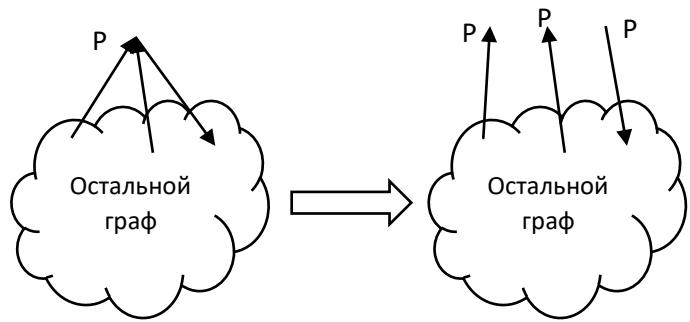

**Рис. 1**

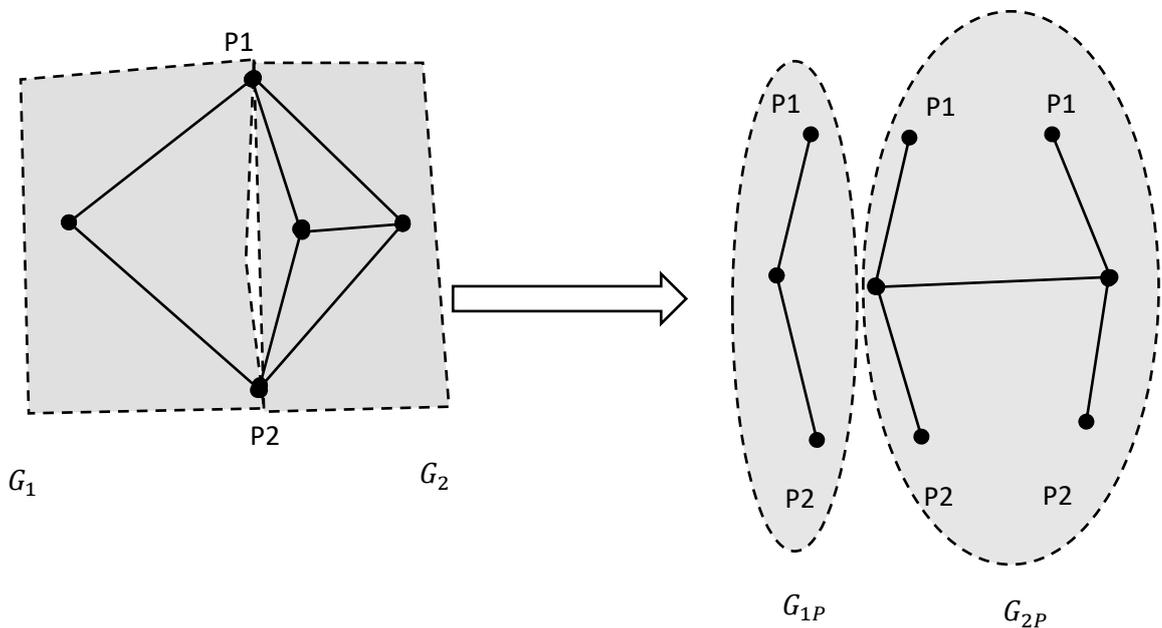

**Рис. 2**

Пусть $G$ - связный граф, на котором задана КЗП. Определим следующее преобразование (назовем его P-приведением, или $Pcut$) – рассмотрим все узлы с заданным давлением, и те из них, которые имеют степень >1, разделим, разъединив сходящиеся в них ветви и задав в их концевых узлах то же давление, что в исходном узле (рис.1). Проделав данную операцию со всеми такими узлами, получим граф $Pcut(G)$, в котором все узлы с заданным давлением висящие, т.е. имеют степень 1 (см. пример на рис.2). Граф $Pcut(G)$ может оказаться несвязным, пусть $G_{iP}$ – связные подграфы, из которых он состоит, а $G_i$ – те подграфы, из которых они получаются (см. пример на рис.2). Будем называть подграфы $G_i$ – P – компонентами графа $G$, а $G_{iP}$ – его P-приведенными компонентами. P-компоненты графа пересекаются друг с другом только по узлам с заданным давлением, в которых они соединены. Будем называть граф $G^*$ P-приведенным, если он связный и $Pcut(G^*) = G^*$.

Очевидно, что решение КЗП для графа $G$ эквивалентно ее решению для всех графов $G_i$ или получающихся из них $G_{iP}$ (все результаты решения КЗП совпадают с результатами решения в соответствующих P-компонентов – кроме притоков в узлах с заданным давлением, которые получаются как сумма соответствующих притоков по тем P-компонентам, где соответствующие узлы содержатся). При этом исходные параметры КЗП влияют на решение лишь локально – то есть изменение заданного в узле притока или заданного в узле давления влияет только в пределах тех P-компонент, в которые данный узел входит. В случае узла с заданным давлением это единственная P-компонента, для узла с заданным давлением их может быть несколько. Иными словами, узлы с заданным давлением делят граф на не влияющие друг на друга фрагменты.

При этом P-приведенные графы имеют следующую структуру (облегчающую анализ) – они либо представляют собой отдельную ветвь с заданными концевыми давлениями, либо связный подграф из узлов с заданными расходами и соединяющими их ветвями, с которым дополнительно соединены висящие узлы с заданным давлением. P-компоненты графа имеют такую же структуру, только в них узлы с заданным давлением могут быть не висящими.

Учитывая все сказанное выше, часто бывает удобно изучать свойства решения КЗП для P-приведенных графах – после чего полученные результаты легко обобщаются на случай произвольного связного графа.

### 3. Единственность и свойства монотонности решения КЗП

Пусть $\Omega \subseteq \mathbb{R}$ - некоторое непустое открытое связное множество, и все функции $\varphi_i$ определены на $\Omega \times \Omega$. В качестве $\Omega$ может выступать $\mathbb{R}$, либо полупрямая, либо открытый интервал (ограниченный нижней и верхней границами давления). Рассмотрим сначала функции $\varphi_i$, удовлетворяющие условиям монотонности (строгому возрастанию по начальному давлению и строгому убыванию по конечному), не будем пока накладывать на них даже условия непрерывности! Оказывается, уже и условий монотонности характеристик ветвей достаточно для обеспечения единственности и свойств монотонности решения КЗП.

Пусть $P \in \Omega^{NV}$ – вектор узловых давлений. Тогда уравнения (1) и (2) полностью определяют расходы по ветвям и притоки в узлах. Изучим, как меняются притоки в узлах при изменении узловых давлений.

<u>Лемма 1 (о притоках в узлах).</u>

Пусть $G$ – связный граф с характеристиками ветвей, удовлетворяющим условиям строгой монотонности, $P^1$ и $P^2$ – векторы узловых давлений, $X^1$, $X^2$, $Q^1$, $Q^2$ – определенные по уравнениям (1) и (2) соответствующие векторы расходов и узловых притоков. Определим множества узлов следующим образом:

$V^+$ - множество тех узлов, для которых $P_i^2 > P_i^1$ (давление увеличилось);

$V^-$ - множество тех узлов, для которых $P_i^2 < P_i^1$ (давление уменьшилось);

$V^0$ - множество тех узлов, для которых $P_i^2 = P_i^1$ (давление не изменилось);

$V^{0+}$ - подмножество тех узлов $V^0$, которые соединены ветвью(ями) с узлом из $V^+$, но не соединены с узлами из $V^-$;

$V^{0-}$ - подмножество тех узлов $V^0$, которые соединены ветвью(ями) с узлом из $V^-$, но не соединены с узлами из $V^+$;

$V^{0\pm}$ - подмножество тех узлов $V^0$, которые соединены ветвями с узлом(ами) и из $V^+$, и из $V^-$;

$V^{00}$ - подмножество тех узлов $V^0$, которые не соединены ветвями с узлами из $V^+$ или $V^-$ (то есть соединены только с узлами из $V^0$).

Тогда:

1. Для любого узла $v \in V^{00}$ $Q^2(v) = Q^1(v)$ (приток не изменился).
2. Для любого узла $v \in V^{0+}$ $Q^2(v) < Q^1(v)$ (приток уменьшился).
3. Для любого узла $v \in V^{0-}$ $Q^2(v) > Q^1(v)$ (приток увеличился).
4. Если $V^+ \neq \emptyset$ и $V^+ \neq V$, то $\sum_{v \in V^+} Q^2(v) > \sum_{v \in V^+} Q^1(v)$ (общий приток вырос).
5. Если $V^- \neq \emptyset$ и $V^- \neq V$, то $\sum_{v \in V^-} Q^2(v) < \sum_{v \in V^-} Q^1(v)$ (общий приток уменьшился).

Заметим, что про притоки в узлах из $V^{0\pm}$ априори ничего сказать нельзя.

Доказательство Леммы 1.

Пусть $v$ – узел графа. Для упрощения дальнейших рассуждений поменяем направления инцидентных ему ветвей так, чтобы они начинались в узле $v$. В соответствии с уравнением (1) приток $Q(v)$ в узле $v$ тогда равен сумме расходов по всем соединенным с ним ветвям.

Если $v \in V^{00}$, то для всех соединенных с $v$ ветвей концевые давления $P_i^2$ совпадают с $P_i^1$, поэтому совпадают и расходы по ветвям, а следовательно $Q^2(v) = Q^1(v)$.

Если $v \in V^{0+}$, то у исходящих из $v$ ветвей начальные давления не меняются ($P^2(v) = P^1(v)$), а давления в конечных узлах $v_L$ либо также не меняются ($P^2(v_L) = P^1(v_L)$), либо увеличиваются ($P^2(v_L) > P^1(v_L)$). В первом случае расход по ветвям не меняется, во втором случае (из-за монотонности характеристики ветви) уменьшится. Поскольку $v \in V^{0+}$, есть не менее одной ветви, по которой расход стал меньше. Следовательно, $Q^2(v) < Q^1(v)$. 3-й пункт леммы доказывается аналогично.

Пусть $V^+ \neq \emptyset$ и $V^+ \neq V$. Сложим уравнения все уравнения Кирхгофа (1), соответствующие узлам из $V^+$. В полученной сумме потоки по ветвям, соединяющим узлы из $V^+$, взаимно компенсируются. Поскольку $V^+ \neq V$ и граф связный, множество ветвей, связывающих узлы из $V^+$ с остальными узлами (из $V \setminus V^+$), не пусто. Для простоты рассуждений переориентируем их, если нужно, так, чтобы начальные узлы все лежали в $V^+$, а конечные в $V \setminus V^+$. Тогда сумма притоков по всем узлам $V^+$ равна сумме расходов по всем ветвям из $V^+$ «вовне» $\sum_{v \in V^+} Q(v) = \sum_{v \in V^+, u \in V \setminus V^+} X(e_{vu})$. На каждой из таких ветвей при переходе от $P^1$ к $P^2$ начальное давление увеличилось, а конечное либо уменьшилось, либо не изменилось. Поскольку характеристики этих ветвей монотонны, на каждой из них расход увеличился, а следовательно, увеличилась и сумма расходов – что и доказывает пункт 4 леммы. Пункт 5 леммы доказывается аналогично.

Теперь мы можем установить свойства решения КЗП, вытекающие из монотонности характеристик ветвей.

Теорема 1 (о единственности решения КЗП).

Пусть $G$ – связный граф с характеристиками ветвей, удовлетворяющими условиям строгой монотонности. Тогда для него решение задачи КЗП единственно.

<u>Доказательство Теоремы 1.</u>

В вырожденном случае, когда $N_P = N_V$, теорема выполнена автоматически.

Рассмотрим случай $0 < N_P < N_V$, когда есть и узлы с заданными давлениями, и узлы с заданными притоками (т.е. $V_P \neq \emptyset$ и $V_Q \neq \emptyset$). Допустим, что для одних и тех же множеств $V_P$ и $V_Q$ и одних и тех же заданных узловых давлениях и притоках есть 2 разных решения КЗП (обозначим их индексами 1 и 2). Тогда эти решения должны отличаться вектором узловых давлений. Применим лемму 1. Поскольку в узлах $V_P$ давления фиксированы, $V^+ \subseteq V_Q$, следовательно $V^+ \neq V$. Аналогично $V^- \subseteq V_Q$, и следовательно $V^- \neq V$. Если $V^+ \neq \emptyset$, то согласно пункту 4 леммы 1 $\sum_{v \in V^+} Q^2(v) > \sum_{v \in V^+} Q^1(v)$. Но это невозможно, так как $V^+ \subseteq V_Q$, а все притоки в узлах $V_Q$ заданы и не меняются. Следовательно, $V^+ = \emptyset$. Аналогичным образом доказывается, что и $V^- = \emptyset$. Но это означает, что $V^0 = V$, то есть узловые давления решений 1 и 2 совпадают, а следовательно совпадают и расходы и узловые притоки. Теорема 1 доказана.

Изучим теперь, как влияет на решение КЗП изменение исходных данных – заданных давлений и узловых притоков.

<u>Теорема 2 (о монотонности решения КЗП по заданным притокам).</u>

Пусть $G$ – связный граф с характеристиками ветвей, удовлетворяющими условиям строгой монотонности, на котором заданы 2 задачи КЗП (задача 1 и 2) с совпадающими непустыми множествами $V_P$ и $V_Q$, совпадающими заданными давлениями $P_{fix}$, а заданные притоки $Q_{fix}$ при переходе от 1 к 2 нигде не уменьшились, а в одном или более узлах увеличились.

Тогда во всех P-компонентах $G$, где есть узлы с увеличившимся заданным притоком, во всех узлах $v \in V_Q$ $P^2(v) > P^1(v)$, а во всех узлах $v \in V_P$ $Q^2(v) < Q^1(v)$. В остальных P-компонентах решение не изменилось.

<u>Доказательство Теоремы 2.</u>

Очевидно, что достаточно доказать теорему для каждого отдельного P-компонента. В тех компонентах, где нет узлов с увеличившимся заданным притоком, исходные данные КЗП не изменились, поэтому в силу Теоремы 1 не изменилось и решение.

Пусть $G^*$ -P-компонент, где есть узлы с увеличившимся расходом. Применим к нему лемму 1.

Для $G^*$ $V_P \neq \emptyset$ и $V_P \subseteq V^0$, следовательно $V^- \neq V$. Тогда согласно пункту 5 леммы 1 сумма притоков по всем узлам $V^-$ должна уменьшится. Но это невозможно - $V^-$ может включать только узлы с заданным притоком, а по каждому из них приток не уменьшился! Следовательно $V^- = \emptyset$. Таким образом, все узлы $G^*$ содержатся в $V^0 \cup V^+$, причем $V^+ \neq \emptyset$ (иначе бы притоки не изменились). Если среди узлов с заданным притоком есть те, где давление не изменилось, то среди них должны быть входящие в $V^{0+}$ - в противном случае подграф $G^*$ из узлов с заданным притоком и соединяющих их ветвей был бы несвязным. Но согласно пункту 2 леммы 1 в таких узлах приток должен был бы уменьшится – что невозможно! Следовательно, все узлы с заданным давлением графа $G^*$ входят в $V^+$, то есть давление в них увеличилось.

Пусть теперь $v$ – узел с заданным давлением графа $G^*$. В графе $G^*$ он связан ветвями с узлами с заданным расходом, и только с ними. Переориентируем их так, чтобы они начинались в $v$. Тогда приток $Q(v)$ в графе $G^*$ равен сумме потоков выходящим из них ребрам. Но для каждого из них давление в начале не изменилось, а в конце увеличилось – следовательно в силу монотонности характеристик ребер поток по каждому из них уменьшился, а следовательно уменьшился и приток $Q(v)$.

Тем самым теорема 2 доказана.

Теорема 3 (<u>о монотонности решения КЗП по заданному давлению</u>)

Пусть $G$ – связный граф с характеристиками ветвей, удовлетворяющими условиям строгой монотонности, на котором заданы 2 задачи КЗП (задача 1 и 2) с совпадающими непустыми множествами $V_P$ и $V_Q$, совпадающими заданными притоками $Q_{fix}$, а заданные давления $P_{fix}$ при переходе от 1 к 2 нигде не уменьшились, а в одном узле $v^+$ увеличились.

Тогда:
1. Во всех Р-компонентах $G$, содержащих узел $v^+$, во всех узлах $v \in V_Q$ $P^2(v) > P^1(v)$, а в остальных Р-компонентах во всех узлах $v \in V_Q$ $P^2(v) = P^1(v)$.
2. Если в $G$ есть другие узлы с заданным давлением, кроме $v^+$, то $Q^2(v^+) > Q^1(v^+)$. В других узлах $v \in V_P$ – в тех из них, кто входит в хотя бы один Р-компонент вместе с $v^+$, $Q^2(v) < Q^1(v)$, в остальных $Q^2(v) = Q^1(v)$.
3. Если в $G$ нет других узлов с заданным давлением, кроме $v^+$, то $Q^2(v^+) = Q^1(v^+)$.

<u>Доказательство теоремы 3.</u>

Достаточно доказать теорему для каждого отдельного Р-компонента. В самом деле, это очевидно для пункта 1, а также для пункта 2 относительно притоков в узлах, отличных от $v^+$. Для узла $v^+$ же это следует из того факта, что если в $G$ есть другие узлы с заданным давлением, то $v^+$ обязательно должен попасть в хотя бы один Р-компонент с каким-либо из них (например с тем узлом, до которого наименьшее число ветвей пути из $v^+$) – иначе граф $G$ был бы несвязным.

Итак, пусть $G^*$ -Р-компонент графа $G$. Если $G^*$ не включает $v^+$, то решение КЗП на $G^*$ не изменится. Рассмотрим случай, когда $G^*$ включает $v^+$.

Для анализа изменения давлений в узлах $v \in V_Q$ графа $G^*$ (если таковые есть) используем лемму 1 для графа $G^*$. Поскольку $v^+ \in V^+$, то $V^- \neq V$. При этом $V^-$ может включать только узлы из $V_Q$, поэтому суммарный приток по всем узлам из $V^-$ уменьшится не может. Следовательно, $V^- = \emptyset$. Узлы с заданным притоком + узел $v^+$ + связывающие их ветви образуют в $G^*$ связный подграф. Если в нем есть узлы и из $V^+$, и из $V^0$, то благодаря связности должен быть и узел, принадлежащий $V^{0+}$, а в нем согласно пункту 1 леммы 1 приток должен был бы уменьшится. Следовательно, все узлы из $V_Q$ графа $G^*$ входят в $V^+$. При этом $V^+$ включает в точности все узлы графа $G^*$ из $V_Q$ + узел $v^+$. Если в $G^*$ есть и другие узлы с заданным давлением, то срабатывает 4-й пункт леммы 1. Но притоки во все остальные узлы из $V^+$ фиксированы, поэтому возрасти должен именно приток в узле $v^+$: $Q^2(v^+) > Q^1(v^+)$. Остальные узлы с заданным давлением в $G^*$ попадают в множество $V^{0+}$, и для них срабатывает 1-й пункт Леммы 1.

Наконец, если в $G^*$ $v^+$ - единственный узел с заданным давлением, то постоянство притока в $v^+$ следует из уравнения баланса (4) для графа $G^*$.

Теорема 3 доказана.

## 4. Непрерывность решения КЗП

Добавим теперь к требованию строгой монотонности функций $\varphi_i$ требование непрерывности. Посмотрим, какие дополнительные свойства решения КЗП это влечет.

Обозначим $Y$ вектор, составленный из исходных данных КЗП – первые $N_P$ компонент - заданные узловые давления $P_{fix}$, остальные $N_Q$ – заданные узловые притоки $Q_{fix}$; $Y \in \Omega^{N_P} \times \mathbb{R}^{N_Q}$. Обозначим через $E$ множество тех $Y$, для которых КЗП имеет решение. Можно ли что-то сказать о нем?

Теорема 4 (о непрерывности и свойствах монотонности решения КЗП).

Пусть $G$ – связный граф с характеристиками ветвей, удовлетворяющими условиям строгой монотонности и непрерывности. Тогда:

1. Множество $E$, на котором КЗП имеет решение, гомеоморфно $\Omega^{N_V}$ (или, что то же самое, $\mathbb{R}^{N_V}$) и, следовательно, не пусто, открыто и связно.
2. Все параметры решения (узловые давления, расходы, притоки) являются непрерывными функциями исходных данных.
3. Решение обладает следующими свойствами монотонности от исходных данных:
    a. Давления во всех узлах из $V_Q$ строго монотонно возрастают при возрастании заданных притоков и заданных давлений в узлах своего P-компонента, и не зависят от других исходных данных.
    b. Притоки во всех узлах из $V_P$ строго монотонно убывают при возрастании заданных притоков в узлах P-компонент, в которые они входят, и не зависят от заданных притоков в других узлах.
    c. Приток в узле из $V_P$:
        i. Строго возрастает при росте давления в том же узле, если в $G$ есть другие узлы с заданным давлением.
        ii. Строго убывает при росте давления в другом узле с заданным давлением из того же P-компонента.
        iii. Не меняется во всех остальных случаях.

Доказательство теоремы 4.

Рассмотрим отображение $\Psi: \Omega^{N_V} \to \Omega^{N_P} \times \mathbb{R}^{N_Q}$, ставящее в соответствие вектору узловых давлений $P$ вектор $Y$, составленный из давлений $P_i$ в узлах из $V_P$ и расходов $Q_i$ в узлах из $V_Q$, рассчитанных из $P$ по уравнениям (1)-(2). Фактически отображение $\Psi$ ставит в соответствие вектору $P$ исходные данные той КЗП, решением которой он является. Поэтому $E = \Psi(\Omega^{N_V})$.

Поскольку функции $\varphi_i$ непрерывны, отображение $\Psi$ также является непрерывным. Поскольку функции $\varphi_i$ строго монотонны, в силу теоремы 1 оно является инъективным. Поэтому в силу теоремы Брауэра об инвариантности области отображение $\Psi$ является гомеоморфизмом, а его образ - множество $E = \Psi(\Omega^{N_V})$ - открыто и гомеоморфно $\Omega^{N_V}$. Это доказывает первый пункт теоремы.

Поскольку $\Psi$ – гомеоморфизм, обратное отображение $\Psi^{-1}$ непрерывно на $E$. Это означает, что узловые давления непрерывно зависят от исходных данных КЗП. Поскольку $\varphi_i$ также непрерывны, рассчитываемые по уравнениям (1) и (2) расходы по ветвям и притоки в узлах также непрерывно зависят от исходных данных КЗП.

3-й пункт теоремы прямо вытекает из теорем 2 и 3.

Пусть $P_{fix}$ задано. Обозначим через $E_{Pfix}$ множество тех $Q_{fix}$, для которых КЗП имеет решение.

Пусть $N_Q > 0$. Рассмотрим отображение $\Psi_{Pfix}: \Omega^{N_Q} \to \mathbb{R}^{N_Q}$, ставящее в соответствие давлениям $P_{var}$ притоки $Q_{fix}$, рассчитанные по уравнениям (1) и (2) по $P_{fix}$ и $P_{var}$. Тогда $E_{Pfix} = \Psi_{Pfix}(\Omega^{N_Q})$. Отображение $\Psi_{Pfix}$ является просто ограничением отображения $\Psi$. Оно также является непрерывным и инъективным, и поэтому является гомеоморфизмом между $\Omega^{N_Q}$ и $E_{Pfix}$.

Пусть $K^{(i)} = \left[P_{min}^{(i)}, P_{max}^{(i)}\right] \subset \Omega$ – замкнутые интервалы ($P_{min}^{(i)} < P_{max}^{(i)}$), $K = \prod_{i=1}^{N_Q} K^{(i)}$. Тогда $E_K = \Psi_{Pfix}(K)$ – множество векторов притоков $Q_{fix}$ для заданного $P_{fix}$, для которых решение КЗП существует и лежит в $K$. Множество $E_K$ гомеоморфно $N_Q$-мерному параллелепипеду $K$. Что оно собой представляет?

$K$ и $E_K$ компактны. $\Psi_{Pfix}$ отображает границу $K$ в границу $E_K$: $\partial E_K = \Psi_{Pfix}(\partial K)$, а внутренность $K$ – во внутренность $E_K$. Граница $\partial K$ представляет собой совокупность $2N_Q$ $N_Q - 1$ – мерных поверхностей (граней) и гомеоморфна $N_Q - 1$ – мерной сфере. В соответствии с теоремой Жордана-Брауэра $\partial E_K$ разбивает $\mathbb{R}^{N_Q}$ на внутреннюю и внешнюю связные компоненты (области). Тогда можно показать, что внутренность $E_K$ в точности совпадает с областью, ограниченной $\partial E_K$ (иначе у $E_K$ появились бы точки границы, не принадлежащие $\partial E_K$) – то есть **вся** область, ограниченная $\partial E_K$, вместе с самой границей, совпадает с $E_K$.

Возникает вопрос – может ли этот небольшой экскурс в общую топологию привести к каким-либо практическим результатам? Оказывается, в определенной степени может.

<u>Теорема 5 (о существовании решения КЗП для промежуточных притоков).</u>

Пусть $G$ – связный граф с характеристиками ветвей, удовлетворяющими условиям строгой монотонности и непрерывности, на котором определена КЗП с $N_Q > 0$.

Тогда если $Q_{fix}^1 \in E_{Pfix}$, $Q_{fix}^2 \in E_{Pfix}$, $Q_{fix}^1 < Q_{fix}^2$ (то есть строгое неравенство выполняется для всех компонент векторов), и $Q_{fix}^1 \leq Q_{fix} \leq Q_{fix}^2$, то $Q_{fix} \in E_{Pfix}$. При этом для соответствующих им векторов узловых давлений $P_{var}^1 \leq P_{var} \leq P_{var}^2$.

Иными словами, если КЗП имеет решение для некоторых «граничных» притоков, то существует решение и для всех их промежуточных значений (интуитивно достаточно ожидаемый результат).

<u>Доказательство теоремы 5.</u>

В соответствии с пунктом 3а теоремы 4 $P_{var}^1 < P_{var}^2$. Примем $P_{min}^{(i)} = P_i^1$, $P_{max}^{(i)} = P_i^2$ и рассмотрим соответствующий $N_Q$ параллелепипед $K_P$ и соответствующее ему множество $E_{KP} = \Psi_{Pfix}(K_P)$, а также параллелепипед $K_Q$, определяемый условиями $Q_{fix}^1 \leq Q_{fix} \leq Q_{fix}^2$. Тогда оказывается, что $K_Q \subseteq E_{KP}$. Для случая $N_Q = 1$ это очевидно следует из пункта 3ci теоремы 4 (в этом случае $K_Q$ и $E_{KP}$ просто совпадающие отрезки). Рассмотрим случай $N_Q > 1$.

В этом случае граница $\partial E_{KP} = \Psi_{Pfix}(\partial K_P)$ состоит из $N_Q$ пар $N_Q - 1$ – мерных гиперповерхностей, являющихся образами граней параллелепипеда $K_P$ при отображении $\Psi_{Pfix}$. Рассмотрим пару гиперповерхностей, являющуюся образом пары граней $\partial K_P$ по i-й компоненте вектора $P_{var}$ – эта пара граней задается условиями $P_{i\,var} = P_{var}^1$ и $P_{i\,var} = P_{var}^2$, и $P_{var}^1 \leq P_{j\,var} \leq P_{var}^2$ для всех $j \neq i$. Для образа любой точки $P_{var}$, принадлежащей «нижней» грани ($P_{i\,var} = P_{var}^1$), при отображении $\Psi_{Pfix}$, справедливо неравенство $Q_i \leq Q_{i\,fix}^1$. Это следует из пунктов 2 и 3 теоремы 3 (надо последовательно

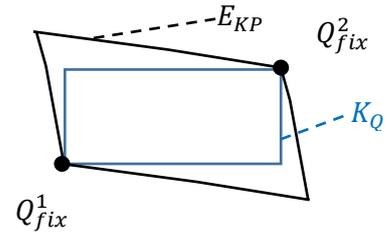

Рис. 3

увеличивать, если требуется, все j-е узловые давления, чтобы перейти от $P_{var}^1$ к $P_{var}$, при этом приток $Q_i$ будет уменьшаться или оставаться постоянным). Аналогичным образом для всех точек «верхней» грани $Q_i \geq Q_{i\,fix}^2$. Тем самым $K_Q$ лежит между этой парой составляющих границу $\partial E_{KP}$ гиперповерхностей. Так как это имеет место для всех составляющих границу $\partial E_{KP}$ пар гиперповерхностей, $K_Q \subseteq E_{KP}$. (Для наглядности на рис.3 схематически показан случай $N_Q = 2$).

Если $Q_{fix}^1 \leq Q_{fix} \leq Q_{fix}^2$, то $Q_{fix} \in K_Q$ и, следовательно, $Q_{fix} \in E_{KP}$, то есть решение КЗП для $Q_{fix}$ существует. Неравенство $P_{var}^1 \leq P_{var} \leq P_{var}^2$ следует из пункта 3а теоремы 4, что и завершает доказательство теоремы 5.

## 5. Теорема о существовании решения КЗП

Теорема 6 (о существовании решения КЗП).
Пусть $G$ - связный граф с характеристиками всех ветвей из множества $\tilde{Z}_a^2$, для которого задана КЗП. Тогда КЗП всегда имеет решение.
Доказательство теоремы 6.
Доказательство проведем по индукции по числу ветвей $N_E$ графа $G$. Фактически это доказательство описывает некоторый рекурсивный алгоритм поиска решения. Конечно, вряд ли он численно эффективен, но тем не менее успешно позволяет установить существование решения.

База индукции ($N_E = 1$). Для графов, состоящих из одной ветви (и 2 узлов), для вырожденного случая, когда задано давление в 2 узлах, КЗП имеет решение по определению. Для случая задания давления в одном узле и притока в другом, решение КЗП существует в силу того, что функции $f_L$ и $f_F$ для ветвей с характеристиками из $\tilde{Z}_a^2$ определены на всей области $\mathbb{R} \times \mathbb{R}$ (см. раздел 2 статьи).

Шаг индукции. Пусть граф $G$ имеет $N_E > 1$ ветвей и существование решения КЗП доказано для всех графов с числом ветвей меньше $N_E$. Рассмотрим граф $G' = Pcut(G)$. Решение задачи КЗП на графе $G$ эквивалентно ее решению на $G'$. Если последний несвязен и содержит несколько связных подграфов (P-приведенных компонент), то каждый из них содержит меньше $N_E$ ветвей, и следовательно КЗП на каждом из них имеет решение, а следовательно и КЗП на $G'$.

Остается случай, когда $G'$ связен, и, следовательно, является P-приведенным. Выберем в $G'$ некоторый узел $v'$ с заданным давлением $P_{fix}(v')$. Этот узел является висящим и соединен

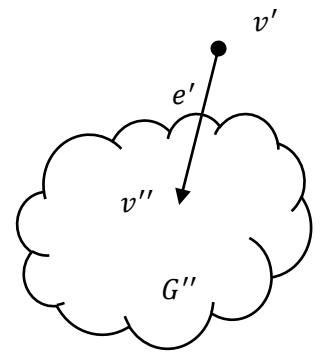

Рис. 4

ветвью $e'$ с некоторым узлом $v''$, являющимся узлом с заданным притоком $Q_{fix}(v'')$. Для упрощения дальнейших рассуждений поменяем, если требуется, направление ветви $e'$, так чтобы она шла из узла $v'$ в узел $v''$. Пусть $G''$ - граф, полученный из $G'$ удалением узла $v'$ и ветви $e'$ (см. рис.4). Он является связным и содержит $N_E - 1$ ветвь, то есть согласно предположению индукции на нем любая КЗП должна иметь решение. Возможны 2 варианта – 1) когда в графе $G'$ есть только один узел с заданным давлением (а именно – узел $v'$); 2) когда в графе $G'$ есть более одного узла с заданным давлением – и, следовательно, такие узлы есть в $G''$.

1-й вариант. В этом случае выполняем «прямой просчет» ветви $e'$. Из уравнения (4) в этом случае вытекает, что для искомого решения $Q(v') = -\sum_{v \in G', v \neq v'} Q_{fix}(v)$. Поскольку узел $v'$ висящий, примем расход на ветви $e'$ равным $X(e') = Q(v')$ и рассчитаем давление в узле $v''$ по функции $f_L$ для ветви $e'$: $P(v'') = f_{e'L}(P_{fix}(v'), X(e'))$. Зададим на графе $G''$ КЗП с теми же условиями, что на $G'$, но вместо притока в узле $v''$ зададим рассчитанное давление. Решение такой задачи на $G''$ существует, и вместе с заданным давлением в узле $v'$ дает решением исходной КЗП на $G'$. Это очевидно для всех узлов, кроме $v''$. В узле $v''$ для найденного решения $Q(v'') = -\sum_{v \in G'', v \neq v''} Q_{fix}(v) - X(e') = -\sum_{v \in G'', v \neq v''} Q_{fix}(v) + \sum_{v \in G', v \neq v'} Q_{fix}(v) = Q_{fix}(v'')$, что и требовалось.

2-й вариант. В этом случае будем искать такое давление $P''$ в узле $v''$, чтобы решение КЗП на $G''$ (которое всегда существует по предположению индукции) с заданным давлением $P''$ и с теми же условия в остальных узлах $G''$, что и на $G'$, в комбинации с расходом по ветви $e'$, соответствующим концевым давлениям $P_{fix}(v')$ и $P''$, обеспечивало решение исходной КЗП на $G'$. Чтобы подобное решение было решением исходной КЗП на $G'$, требует только обеспечить заданный приток в узле $v''$. Соответствующее условие записывается в виде
$$Q_{G''var}(v'', P'') - \varphi_{e'}(P_{fix}(v'), P'') = Q_{fix}(v'')$$
где $\varphi_{e'}$ - функция $\varphi$ – для ветви $e'$, $Q_{G''var}(v'', P'')$ - приток в узле $v''$ для решения КЗП на $G''$ с заданным давлением $P''$ в узле $v''$ и теми же, что на $G'$, условиями в остальных узлах $G''$.

Определим функцию $q(P'')$ следующим образом:
$$q(P'') = Q_{G''var}(v'', P'') - \varphi_{e'}(P_{fix}(v'), P'') - Q_{fix}(v'')$$
В соответствии с условиями на характеристики ветвей $-\varphi_{e'}(P_{fix}(v'), P'')$ как функция от $P''$ принадлежит множеству $\tilde{Z}_a$. Согласно предположению индукции $Q_{G''var}(v'', P'')$ как функция от $P''$ определена на всем $\mathbb{R}$. Согласно пункту 2 теоремы 4 она непрерывна по $P''$, а согласно пункту 3ci теоремы 4 строго возрастает. Следовательно, функция $q(P'')$ также принадлежит множеству $\tilde{Z}_a$, а значит принимает в некоторой точке нулевое значение. Эта точка и будет искомой величиной давления $P''$.

Теорема 6 имеет общий характер. Однако для практических целей, как уже было сказано, желательно установить условия существования решения для гидравлических цепей с ограниченной областью определения замыкающих соотношений по давлению, и без «асимптотических» условий 5), 6). Это в определенных границах представляется возможным, и автор планирует осветить эту тему в отдельной статье – здесь же приведем наиболее простой (но очень распространенный и важный) частный случай.

Будем называть гидравлическую цепь пассивной, если все ее ветви пассивные. Частный случай КЗП, в которой заданные притоки $Q_{fix} = 0$, будем называть задачей расчета пропускной способности (ЗРПС).

Теорема 7 (о существовании решения задачи ЗРПС для пассивных цепей)

Пусть $G$ – связный граф пассивной гидравлической цепи с характеристиками ветвей, определенными на $\Omega \times \Omega$ (где $\Omega \subseteq \mathbb{R}$ - некоторое непустое открытое связное множество), удовлетворяющими условиям строгой монотонности и непрерывности. Пусть на $G$ задана КЗП с $Q_{fix} = 0$ и $P_{fix} \in \Omega^{N_P}$. Тогда решение КЗП существует, причем для него $P_{var} \in [\min(P_{fix}), \max(P_{fix})]^{N_Q}$.

Доказательство теоремы 7.

Доказательство проводится полностью аналогично доказательству теоремы 6, с небольшой модификацией ключевых шагов.

Для базы индукции в случае заданного на одной ветви одного узлового давления и нулевого притока в другом узле решением будет то же давление в другом узле.

Аналогично, для 1-го варианта для шага индукции можно просто присвоить давление $P_{fix}(v')$ всем узлам $G''$.

Для 2-го варианта функция $q(P'')$ запишется в виде $q(P'') = Q_{G''\,var}(v'', P'') - \varphi_{e'}(P_{fix}(v'), P'')$, и ее нуль следует искать на отрезке между давлением $P^{(1)} = P_{fix}(v')$ и давлением $P^{(2)} = P_{G''\,var}(v')$ решения задачи КЗП с на графе $G''$ с теми же условиями в узлах, что и в исходной задаче. При этом $q(P^{(1)}) = Q_{G''\,var}(v'', P^{(1)})$ и $q(P^{(2)}) = -\varphi_{e'}(P^{(1)}, P^{(2)})$. Если $P^{(1)} = P^{(2)}$, то это и есть нуль функции $q(P'')$. Если $P^{(1)} > P^{(2)}$, то $q(P^{(2)}) < 0$, а $q(P^{(1)}) > 0$ (из-за строгой монотонности $\varphi_{e'}$ и пункту 3сi теоремы 4), а если $P^{(1)} < P^{(2)}$ – то наоборот, так что нуль $q(P'')$ в любом случае лежит на отрезке $[P^{(1)}, P^{(2)}]$. Осталось заметить, что оба этих значения лежат на отрезке $[\min(P_{fix}), \max(P_{fix})]$ (значение $P^{(2)}$ – в силу предположения индукции).

## 6. Матрицы чувствительности и свойства матрицы Максвелла

Рассмотрим теперь ситуацию, когда замыкающие соотношения ветвей непрерывно дифференцируемы в окрестности решения КЗП, выведем важнейшие матрицы чувствительности решения к исходным данным (аналогичные полученным в [16, 17, 18]), и посмотрим, как их свойства соответствуют установленным законам монотонности решения КЗП.

Обозначим $d_{Fi} = \partial \varphi_i(P_F, P_L)/\partial P_F$, $d_{Li} = -\partial \varphi_i(P_F, P_L)/\partial P_L$. Тогда в силу монотонности $\varphi_i$ справедливы неравенства $d_{Fi} \geq 0$ и $d_{Li} \geq 0$. Далее мы будем рассматривать только «невырожденный» случай, когда производные отличны от нуля, т.е. $d_{Fi} > 0$ и $d_{Li} > 0$.

Определим диагональные матрицы $D_F$ и $D_L$ с $d_{Fi}$ и $d_{Li}$ на диагонали. Тогда из уравнений (3) и (2) получаем

$$dX = (D_F A_F^T + D_L A_L^T)dP \qquad (5)$$
$$dQ = A(D_F A_F^T + D_L A_L^T)dP \qquad (6)$$

Очевидно, что для вырожденного случая $N_Q = 0$ решение непрерывно дифференцируемо. Рассмотрим случай $N_Q > 0$. Получим Якобиан отображения $\Psi_{P_{fix}}$. Для этого перенумеруем узлы графа так, чтобы сначала шли узлы с из $V_Q$ (с заданным расходом), а затем из $V_P$ (с заданным давлением) и разобьем вектора и матрицы на соответствующие блоки:

$$P = \begin{pmatrix} P_{var} \\ P_{fix} \end{pmatrix}, Q = \begin{pmatrix} Q_{fix} \\ Q_{var} \end{pmatrix}, A = \begin{pmatrix} A_Q \\ A_P \end{pmatrix}, A_F = \begin{pmatrix} A_{FQ} \\ A_{FP} \end{pmatrix}, A_L = \begin{pmatrix} A_{LQ} \\ A_{LP} \end{pmatrix} \quad (7)$$

Тогда уравнения (5) и (6) запишутся в виде

$$dX = (D_F A_{FQ}^T + D_L A_{LQ}^T) dP_{var} + (D_F A_{FP}^T + D_L A_{LP}^T) dP_{fix} \quad (8)$$
$$dQ_{fix} = A_Q (D_F A_{FQ}^T + D_L A_{LQ}^T) dP_{var} + A_Q (D_F A_{FP}^T + D_L A_{LP}^T) dP_{fix} \quad (9)$$
$$dQ_{var} = A_P (D_F A_{FQ}^T + D_L A_{LQ}^T) dP_{var} + A_P (D_F A_{FP}^T + D_L A_{LP}^T) dP_{fix} \quad (10)$$

Матрица $\widetilde{M} = A_Q (D_F A_{FQ}^T + D_L A_{LQ}^T)$, связывающая $dQ_{fix}$ и $dP_{var}$ (называемая также модифицированной матрицей Максвелла [11, 13]), и есть матрица Якоби отображения $\Psi_{Pfix}$. Эта матрица обладает целым рядом замечательных свойств, которые мы далее обсудим. Пока лишь отметим, что для связного графа и $d_{Fi} > 0$, $d_{Li} > 0$ матрица $\widetilde{M}$ всегда не вырождена. Из этого, в соответствии с теоремой о неявной функции, следует, что все параметры решения КЗП являются локально непрерывно дифференцируемыми функциями исходных данных. Обозначим

$$\widetilde{M}_{PP} = A_P (D_F A_{FP}^T + D_L A_{LP}^T), \widetilde{M}_{PQ} = A_P (D_F A_{FQ}^T + D_L A_{LQ}^T), \widetilde{M}_{QP} = A_Q (D_F A_{FP}^T + D_L A_{LP}^T) \quad (11)$$

Тогда, выражая из (9), (10) изменение параметров решения через изменение исходных данных, получим

$$dP_{var} = \widetilde{M}^{-1} dQ_{fix} - \widetilde{M}^{-1} \widetilde{M}_{QP} dP_{fix} \quad (12)$$
$$dQ_{var} = \widetilde{M}_{PQ} \widetilde{M}^{-1} dQ_{fix} + (\widetilde{M}_{PP} - \widetilde{M}_{PQ} \widetilde{M}^{-1} \widetilde{M}_{QP}) dP_{fix} \quad (13)$$

Уравнения (12), (13) дают основные матрицы чувствительности нашей задачи. Для «традиционных» цепей они совпадают с полученными в [16, 17, 18].

Изучим теперь свойства полученных матриц, прежде всего – модифицированной матрицы Максвелла. Запишем последнюю в виде

$$\widetilde{M} = A_Q (D_F A_{FQ}^T + D_L A_{LQ}^T) = A_{FQ} D_F A_{FQ}^T + A_{LQ} D_L A_{LQ}^T + A_{LQ} D_F A_{FQ}^T + A_{FQ} D_L A_{LQ}^T$$

Матрицы $A_{FQ} D_F A_{FQ}^T$ и $A_{LQ} D_L A_{LQ}^T$ – диагональные. Первая содержит в i-й ячейке диагонали (соответствующей i-му узлу $V_Q$) сумму $d_F$ всех выходящих из узла ветвей вторая – сумму $d_L$ всех входящих в узел ветвей. Матрица $A_{FQ} D_L A_{LQ}^T$ в i-й строке и j-м столбце содержит сумму $-d_L$ всех ветвей, выходящих из узла i и входящих в узел j, а матрица $A_{LQ} D_F A_{FQ}^T$ – сумму $-d_F$ всех ветвей, входящих в узел i и выходящих из j. В итоге матрица $\widetilde{M}$ содержит на диагонали сумму $d_F$ всех выходящих из узла и сумму $d_L$ всех входящих в узел ветвей (кроме ветвей-петель); недиагональный элемент $m_{ij}$ содержит сумму величин для всех соединяющих вершины i и j ветвей: $-d_L$ для ветвей из i-го узла в j-й и $-d_F$ для ветвей из j-го узла в i-й.

Матрица $\widetilde{M}_{PP}$ имеет такую же структуру, как $\widetilde{M}$, но только для узлов $V_P$.

Для матриц $\widetilde{M}_{QP}$ и $\widetilde{M}_{PQ}$ имеем $\widetilde{M}_{QP} = A_{FQ} D_F A_{FP}^T + A_{LQ} D_L A_{LP}^T + A_{FQ} D_L A_{LP}^T + A_{LQ} D_F A_{FP}^T$ и $\widetilde{M}_{PQ} = A_{FP} D_F A_{FQ}^T + A_{LP} D_L A_{LQ}^T + A_{FP} D_L A_{LQ}^T + A_{LP} D_F A_{FQ}^T$. Первые 2 слагаемых в обоих выражениях равны нулю, и в итоге обе матрицы содержат в ячейке ij сумму величин, соединяющих узлы i и j: $-d_L$ для ветвей из i-го узла в j-й и $-d_F$ для ветвей из j-го узла в i-й. При этом $\widetilde{M}_{PQ} = \widetilde{M}_{QP}^T$

Матрица $\widetilde{M}$ для цепей с зависящими от давления замыкающими соотношения, не является симметричной (как для «традиционных» цепей). Тем не менее, она сохраняет многие свои замечательные свойства.

Прежде всего, $\widetilde{M}$ является матрицей со слабым диагональным преобладанием по столбцам. При этом строгое диагональное преобладание имеет место в столбцах, соответствующим узлам, соединенным ветвями с узлами из $V_P$ (которые дают дополнительный вклад в диагональные элементы). Поскольку граф связен и $N_P > 1$, из любого узла $V_Q$ существует путь до какого-либо узла из $V_P$ с заданным давлением, а следовательно и до узла из $V_Q$, соединенного с $V_P$, причем ветвям пути соответствуют ненулевые элементы вне диагонали. Это делает матрицу $\widetilde{M}$ принадлежащей классу WCDD (Weakly Chained Diagonally Dominant), а следовательно, невырожденной (в российской литературе на этот факт принято ссылаться как на обобщение теоремы Ольги Тауски [19]).

Однако это еще не все замечательные свойства матрицы $\widetilde{M}$. Одновременно она является L-матрицей, то есть матрицей, чьи диагональные элементы положительны, а не диагональные элементы отрицательны или равны нулю. Оказывается, уже давно известно ([20]), что все матрицы, одновременно относящие к классам WCDD и L, являются невырожденными M-матрицами (матрицами с неположительными не диагональными элементами, действительная часть собственных значений которых положительна). Этот класс матриц привлекает в последние годы большое внимание и возникает в разнообразных областях математики (дифференциальных уравнениях, марковских цепях и др. [21]). В частности, для них установлено много полезных оценок для их определителя и различных норм (см., например [22, 23]). M-матрицы монотонны – в частности, обратные к ним матрицы содержат только неотрицательные элементы.

Изучим более детально свойства и структуру обратной модифицированной матрицы Максвелла $\widetilde{M}^{-1}$.

Невырожденная M-матрица $\widetilde{M}$ с учетом теоремы Перрона-Фробениуса может быть представлена в виде [21] $\widetilde{M} = sI - B$, где $I$ – единичная матрица, матрица $B$ неотрицательна ($B \geq 0$), а $s$ больше спектрального радиуса матрицы $B$ ($s > \rho(B)$). Тогда обратная матрица может быть представлена в виде

$$\widetilde{M}^{-1} = s^{-1}\left[I + \sum_{i=1}^{\infty} s^{-i} B^i\right] \tag{14}$$

При этом ряд в (14) сходится в силу неравенства $\rho(s^{-1}B) < 1$.

Очевидно, что все элементы матрицы $\widetilde{M}^{-1}$ неотрицательны, причем все диагональные элементы положительны. Какие именно не диагональные элементы положительны? Очевидно, в матрице $B$ положительны те, и только те не диагональные элементы, которые отрицательны в матрице $\widetilde{M}$ – то есть соответствующие узлам, соединенным ветвью. В матрице $B^i$ положительны все не диагональные элементы, соответствующие узлам, соединенным путем из i ветвей со всеми узлами в $V_Q$ (а, также, возможно, и часть элементов, соединенных путем из меньшего количества ветвей – это зависит от диагональных элементов матрицы $B$). В итоге из формулы (14) получаем, что в матрице $\widetilde{M}^{-1}$ положительны те, и только те элементы, которые соответствуют паре узлов, соединенных некоторым путем со всеми узлами в $V_Q$.

Перенумеруем узлы $V_Q$ таким образом, чтобы узлы из каждого P-компонента шли подряд. Тогда матрицы $\widetilde{M}$ и $\widetilde{M}^{-1}$ примут вид блочно-диагональных – каждый блок соответствует невырожденному (содержащему узлы из $V_Q$) P-компоненту. Поскольку узлы $V_Q$ в каждом P-компоненте образуют связный подграф, тем самым доказана следующая теорема.

<u>Теорема 8 (об обратной матрице Максвелла)</u>

Обратная модифицированная матрица Максвелла $\widetilde{M}^{-1}$ имеет блочно-диагональную структуру, соответствующую невырожденным P-компонентам графа, причем все элементы каждого блока положительны. Если граф содержит только 1 P-компонент, то все элементы матрицы $\widetilde{M}^{-1}$ положительны.

Теперь мы может установить соответствие между свойствами матриц чувствительности. Установленный теоремой 8 результат, очевидно, соответствует п.3а теоремы 4 в части чувствительности $dP_{var}$ от $dQ_{fix}$. Матрица $-\widetilde{M}^{-1}\widetilde{M}_{QP}$, определяющая чувствительность $dP_{var}$ от $dP_{fix}$, имеет только положительные элементы для пар узлов, принадлежащих одному P-компоненту, и нулевые остальные. Это также соответствует п.3а теоремы 4. Матрица $\widetilde{M}_{PQ}\widetilde{M}^{-1}$, определяющая чувствительность $dQ_{var}$ к $dQ_{fix}$, имеет отрицательные элементы для принадлежащих одному P-компоненту узлов, и нулевые остальные. Это соответствует п.3b теоремы 4.

Наконец, рассмотрим матрицу $\widetilde{M}_{PP} - \widetilde{M}_{PQ}\widetilde{M}^{-1}\widetilde{M}_{QP}$, определяющую чувствительность $dQ_{var}$ к $dP_{fix}$. Заметим, что в силу уравнения (4) сумма компонентов $dQ_{var}$ при $dQ_{fix} = 0$ должна быть равна нулю. Это означает, что сумма строк матрицы $\widetilde{M}_{PP} - \widetilde{M}_{PQ}\widetilde{M}^{-1}\widetilde{M}_{QP}$ равна нулю. Для случая, когда узел с заданным давлением только один, вся матрица чувствительности равна нулю. Для вырожденных P-компонент диагональные элементы положительные, а не диагональные отрицательные. Наконец, для невырожденных P-компонент все не диагональные элементы матрицы отрицательны – поэтому из-за равенства нулю строк все диагональные элементы положительны. Это соответствует п.3с теоремы 4.

**Приложение 1. Формулировка теоремы о неявной функции для непрерывных функций**

Пусть отображение $F$ из $\mathbb{R}^n \times \mathbb{R}^m$ в $\mathbb{R}^m$ определено в некоторой окрестности $W$ точки $(x_0, y_0) \in \mathbb{R}^n \times \mathbb{R}^m$ и удовлетворяет следующим условиям:

- $F$ является непрерывным в $W$
- $F(x_0, y_0) = 0$
- Существуют окрестности $U$ и $V$ точек $x_0$ и $y_0$ в $\mathbb{R}^n$ и $\mathbb{R}^m$ соответственно, такие что $U \times V \subseteq W$, и для любого $x \in U$ отображение $y \to F(x, y)$ является локально взаимно-однозначным

Тогда существует такое непрерывное отображение $f: U \to V$,
что $F(x, y) = 0 \Leftrightarrow y = f(x)$ для всех $x \in U$ и $y \in V$.